\documentclass{article}
\begin{document}
\begin{center}
{\Large Casimir Effect in Hyperbolic Polygons }
\end{center}
\vspace{5mm}
\begin{center}
H. Ahmedov\footnote{E-mail: hagi@gursey.gov.tr}
\begin{flushleft}
Feza Gursey Institute, P.O. Box 6, 81220, \c{C}engelk\"{o}y,
Istanbul, Turkey.
\end{flushleft}

\end{center}

\begin{center}
{\bf Abstract}
\end{center}
We derive  a  trace  formula for  the spectra of quantum mechanical
systems in hyperbolic polygons  which are the fundamental domains of
discrete isometry groups acting in the two dimensional hyperboloid.
Using this trace formula and the point splitting regularization
method we calculate the Casimir energy for a  scalar fields in such
domains. The   dependence of the vacuum energy on the number of
vertexes is established.

\vspace{1cm} \noindent {\bf 1. Introduction}
\\
\\
Sign of the Casimir energy is known to be dependent on the
dimension, topology, metric properties of manifolds  and on the
shape of boundaries where fields under consideration vanishes
\cite{Bor}. For example the vacuum energy in the three dimensional
ball is positive \cite{ball} whereas in the original parallel plane
configuration it is negative \cite{PP}. No general approach is known
to investigate the dependence of the Casimir energy  on the geometry
of boundaries for compact domains. Technical difficulties related to
the solving of the Dirichlet problem restrict the number of cavities
for which explicit results can be obtained. It seems that only
those cavities which have additional symmetries ( such as the
rotational symmetry for the ball ) are treatable. Cavities which are
the fundamental domains of discrete isometry groups give us another
class of solvable physical system: Symmetries of isometry discrete
groups can be used to find spectra of  such systems. However the
Euclidean spaces admits only finite number of crystallographic
groups and  consequently only special cavities with fixed shapes can
be treated  by the reflection method \cite{HA}.
\\
\\
Hyperbolic spaces in contrast to the Euclidean ones admit infinite
number different cavities which can be treated by the reflection
method \cite{Fom}.  This may give us the possibility to find the
analytic expression for the Casimir energy as a function of the
boundary configuration. We restrict our consideration on the two
dimensional hyperbolic space $H^2$. We consider polygons bounded by
an arbitrary number geodesics ( the analogue of polygons in the flat
case ). Duality between the configuration and momentum spaces are
not well known for hyperbolic spaces: No individuals formulas for
eigenvalues of the Laplace-Beltrami operator on such domains are
known. The Selberg's trace formula is currently the only available
tool to analyze the sums over the spectra of quantum mechanical
systems on hyperbolic manifolds. This formula manifests the duality
between spectra of compact manifolds and geodesics - elements of
discrete hyperbolic groups which are the fundamental groups for
these manifolds \cite{Sel}. The Selberg's trace formula for
hyperbolic groups ( no points in $H^2$ fixed under the action of the
group ) in the two dimensional hyperboloid $H^2$ is given in
\cite{Ran} and describe the spectra of oriented  two dimensional
manifolds without boundaries. Fundamental groups of the hyperbolic
polygons which we consider are not hyperbolic. To our knowledge no
explicit construction of a trace formula for such groups is known.
\\
\\
The purpose of this paper is to establish the dependence of the
Casimir energy on the boundary configuration for  M-polygons $X$ in
$H^2$ ( polygons with M vertexes ). We derive the trace formula for
the spectra of the Laplace-Beltrami operator in $X$. The trace
formula we obtain is the generalization of the Selberg's trace
formula which includes also boundary effects. Using this trace
formula  and the point splitting regularization method we calculate
the Casimir energy for a minimally coupled massless scalar field.
The explicit formula which defines the dependence of the Casimir
energy on the number of vertices is established. Namely at
$M\rightarrow \infty $ we have
\begin{equation}
    E=\frac{C_0}{R}M\ln M
\end{equation}
where $C_0$ is the positive number given by (\ref{const}).
\\
\\
This paper is arranged as follows. In Section 2 we construct a
Selberg's trace formula for M-polygons in the hyperboloid $H^2$.
Section 3 is devoted to the calculation of the Casimir energy for a
massless scalar field in M-polygons. In Appendix we give an integral
transformation which we use to calculate the Selberg's trace
formula.

\vspace{2cm} \noindent {\bf 2. The Selberg's trace  for hyperbolic
polygons}
\\
\\
The two dimensional hyperboloid $H^2$ in the Lobachevski realization
is the upper half plane $Im (z)> 0$ equipped  with the metric
\begin{equation}\label{MP}
    ds^2=\frac{(dx)^2+(dy)^2}{y^2}, \ \ \    z=x+iy.
\end{equation}
Geodesics in $H^2$ are circles centered at the real axis  and lines
parallel to the imaginary axis. We may parameterize a geodesic by
two points on the extended real line which is the boundary of $H^2$.
By $(x, x^\prime )$  we denote  the circle centered at the real line
and passing through the points $x$ and $x^\prime $. By $(x, \infty
)$  we denote  the line parallel to the imaginary axis and passing
through the points $x$  on the real line.
\\
Consider a polygon  $X$ between M geodesics $L_1, \ L_2, \ \dots,
L_M$. Let $q_j$ be the reflection with respect to the geodesic $L_j$
and $\Gamma$ be the group generated by the reflections $q_1, \ q_2,
\dots, q_M$. We require that $X$  is the fundamental domain of the
discrete group $\Gamma$.
\\
\\
From the Green function
\begin{equation}\label{FGF}
    G_\rho(z,w)=-\frac{1}{2\pi \sqrt{2}} \int_d^\infty dy
    \frac{e^{-i\rho y}}{\sqrt{\cosh y -\cosh d}}
\end{equation}
on the two dimensional hyperboloid \cite{Cam} we construct the one
on $X$ which satisfies the Dirichlet boundary conditions. Here
$d=d(z,w)$  is the invariant distance given by
\begin{equation}\label{dist}
    \cosh d(z.w)=1+\frac{\mid z-w\mid^2}{2Im z Im w}.
\end{equation}
The desired Green  function is
\begin{equation}\label{GF}
\bar{G}_\rho( z,w)=\sum_{\gamma \in \Gamma} \chi (\gamma) G_\rho
(\gamma z,w )
\end{equation}
where
\begin{equation}
     \chi (q_j)=-1
\end{equation}
is the one dimensional representation of $\Gamma$. It satisfies the
equation
\begin{equation}\label{equation}
   [  L+\rho^2+\frac{1}{4} ]\bar{G}_\rho (z,w)=  y^2 \delta (x-x^\prime)\delta
   (y-y^\prime )
\end{equation}
where
 \begin{equation}\label{lapl}
    L=y^2(\partial^2_x+\partial^2_y)
 \end{equation}
is the Laplace-Beltrami operator on  $H^2$. From (\ref{GF}) we
observe the property
\begin{equation}
\bar{G}_\rho( q_jz,w)=- \bar{G}_\rho( z,w).
\end{equation}
From the other side the reflection $q_j$ leaves the geodesic $L_j$
fixed. Thus  the Green function  vanishes on the boundary of $X$.
\\
\\
Consider a two point function
\begin{equation}
    \bar{K}(z.w)=\frac{1}{i\pi}\int_{-\infty}^\infty d\rho \rho h(\rho )\bar{G}_\rho( z,w)
\end{equation}
where  $h(\rho)$ is a function such that the integral is well
defined. The spectral representation
\begin{equation}
    \bar{G}_\rho( z,w)=\sum_{n=0}^\infty \frac{\Psi_n(z)\overline{\Psi_n(w)}}{\rho^2 -\rho^2_n}
\end{equation}
for the Green function  gives  the trace formula
\begin{equation}
   Z[h]= \sum_{n=0}^\infty h(\rho_n) = \int_X d\mu \bar{K}(z.w).
\end{equation}
From the other side  the geodesic representation (\ref{GF}) for the
Green function implies
\begin{equation}
    Z[h]= \sum_{\gamma \in \Gamma}\chi (\gamma) \int_X d\mu K (\gamma
    z,w)
\end{equation}
 The Selberg's trace formula manifests the equality between the
sum over spectra and the sum over geodesics. Here  $\{
\rho^2_n+\frac{1}{4}\}_{n=0}^\infty$ and $\{
\Psi_k(\vec{x})\}_{n=0}^\infty$ are the set of eigenvalues and
eigenfunctions of the Laplacian $-L$,
\begin{equation}
    d\mu = \frac{dxdy}{y^2}
\end{equation}
is the invariant measure on the hyperboloid $H^2$ and
\begin{equation}\label{Kfunction}
  K(z.w)=\frac{1}{i\pi}\int_{-\infty}^\infty d\rho\rho h(\rho )G_\rho(
  z,w)
\end{equation}
with $G_\rho(z,w)$ being the free Green function (\ref{GF}).
\\
\\
Let $H$ be the set of equivalent classes  in $\Gamma$ defined by the
relation:  one identifies $\gamma$ and $\gamma^\prime$ if there
exist $\tilde{\gamma}\in \Gamma$ such that
\begin{equation}
  \gamma^\prime = \tilde{\gamma} \gamma \tilde{\gamma}^{-1}.
\end{equation}
The summation over $\Gamma$ can be replaced by the summation over
the conjugacy classes $H$
\begin{equation}
    Z[h]= \sum_{\gamma \in H}\chi (\gamma) \int_{X/\Sigma_\gamma} d\mu K (\gamma  z,w)
\end{equation}
where
\begin{equation}
    \Sigma_{\gamma}=\{ \tilde{\gamma} \in \Gamma : \ \   \gamma \tilde{\gamma}=
  \tilde{  \gamma } \gamma  \}
\end{equation}
is the  stability group of  $\gamma\in H$. The construction of the
trace formula  is reduced to the problem of the classification of
conjugacy classes and their stability groups.
\\
\\
The stability group of the unit element is the whole group $\Gamma$
and the contribution to the trace formula is \cite{Mark}
\begin{equation}\label{FREE}
    Z_0[h] = \frac{S}{4\pi}\int_{-\infty}^\infty d\rho \rho \tanh\pi\rho h(\rho )
\end{equation}
where $S$ is the area of $X$.
\\
\\
Let  the intersection point $V_j$ of  geodesics $L_j$ and $L_{j+1}$
has the period $P_j$, that is the angle between these geodesics is
$\frac{2\pi}{P_j}$. We assume that $P_j$ is an even number. The
reflections $q_j$ and $q_{j+1}$ generate a finite group $G$ which
lives the vertex $V_j$ fixed. Every geodesic in $H^2$ can be
obtained from the imaginary axis
\begin{equation}
   L_0=(0, \infty )
\end{equation}
by a conformal transformation
\begin{equation}\label{con}
    \gamma z=\frac{az+b}{cz+d}, \ \ \  ad-bc=1
\end{equation}
where $a, b, c, d$ are real numbers. By transforming the geodesic
$L_j$ into $L_0$ and the vertex $V_j$ into the point $z=i$ the
reflections $q_j$ and $q_{j+1}$ take the following form
\begin{equation}
   q_j=q_0, \ \ \ q_{j+1}=k q_0
\end{equation}
where
\begin{equation}\label{reflection}
    q_0 z=-\bar{z}
\end{equation}
is the reflection with respect to the geodesic $L_0$ and
\begin{equation}
    k=\left(
                \begin{array}{cc}
                  \cos \frac{2\pi}{P_j} & \sin\frac{2\pi}{P_j} \\
                  -\sin\frac{2\pi}{P_j} & \cos \frac{2\pi}{P_j} \\
                \end{array}
              \right)
\end{equation}
is the rotation around the point $z=i$ of the order $\frac{P_j}{2}$.
Note that since we are dealing with projective transformations we
have $k^{\frac{P_j}{2}}=-1\equiv 1$.  $G$ is the Dihedral group of
the order $P_j$. The conjugacy classes in $G$ are
\begin{equation}\label{REF}
   \{ q_0, \ kq_0 \}
\end{equation}
and
\begin{equation}
    \{ k, \ k^2, \ \dots k^{\frac{P_j}{4}} \}
\end{equation}
if $\frac{P_j}{2}$ is even and
\begin{equation}
   \{  k, \ k^2, \ \dots  k^{\frac{P_j-2}{4}} \}
\end{equation}
if $\frac{P_j}{2}$ is odd. For the latter case $\Sigma_{k^m}$ is the
cyclic subgroup $G_0$ of $G$ generated by $k$. For the even case we
have  $\Sigma_{k^m}=G_0$ if $m\neq \frac{P_j}{4}$ and
$\Sigma_{k^m}=G$ if $m= \frac{P_j}{4}$. The contribution from the
vertex $V_j$ to the trace formula is ( see (\ref{A} ))
\begin{equation}
    Z_{V_j}[h]=\frac{1}{P_j}\sum_{n=1}^{\frac{P_j}{2}-1}
   \int_0^\infty \frac{dy g(y) \cosh y }{\sinh^2
y+\sin^2\frac{2\pi n}{P_j}}
\end{equation}
where
\begin{equation}\label{gfunction}
   g(y)=\frac{1}{2\pi}\int_{-\infty}^\infty d\rho h(\rho ) e^{-i\rho y}.
\end{equation}
The total contribution to $Z[h]$ from the vertexes of $X$  is
\begin{equation}\label{VERTEX}
    Z_V[h]=\sum_{j=1}^M Z_{V_j}(s).
\end{equation}
Let $\Lambda$ be the set of all geodesic on $H^2$ which can be
obtained from the geodesics $L_j$, $j=1,\dots, M$ which form the
boundary of $X$ by the action of the group $\Gamma$.  We consider a
class of fundamental domains $X$ with the following property: for
any nonintersecting pair of geodesics  $L$ and $L^\prime$ in
$\Lambda$ there exist a geodesic $\tilde{L}$ in $\Lambda$ which
intersects $L$ and $L^\prime$ at the right angles. \footnote{It
seems that this property is valid for any discrete group $\Gamma$
generated by reflections and having nonzero compact fundamental
domain in $H^2$, but the proof for general case appears to be
difficult } If $q$ and $q^\prime$ are the reflections with respect
to the geodesics $L$ and $L^\prime$ then $\gamma=qq^\prime$ is the
translation along the geodesic $\tilde{L}$, that is $\gamma$
transforms this geodesic into itself. Thus for any geodesic in
$\Lambda$ there exists a translation along this geodesic.
\\
\\
Now we are ready to consider the reflections (\ref{REF}).
$\Sigma_{q_0}$ is generated by the reflection $q_0$ and by the
translation
\begin{equation}\label{trans}
      \gamma =\left(
                \begin{array}{cc}
                  e^{\frac{l_\gamma}{2}} & 0 \\
                  0 & e^{-\frac{l_\gamma}{2}} \\
                \end{array}
              \right)
\end{equation}
along the imaginary axis. Here $l_\gamma$ is the length of $\gamma$.
The fundamental domain $\Omega$ of $\Sigma_{q_0}$ is the half of the
strip between two circles of radii 1 and $e^{l_\gamma}$ centered at
$z=0$:
\begin{equation}\label{FUN}
    \Omega =\{z\in H^2: 1< \mid z\mid < e^{l_\gamma}, \ \  Re z>0
    \}.
\end{equation}
All other edges of $X$ can be treated in a similar fashion. The
contribution to the trace formula from the geodesic which form the
boundary of $X$  is ( see (\ref{B}))
\begin{equation}\label{GEODESIC}
    Z_{L}[h]=\frac{ g(0)}{4}\sum_{j=1}^M l_{\gamma_j}
\end{equation}
where $\gamma_j$ is the length of the  translation along the
geodesic $L_j$.
\\
\\
Now we consider hyperbolic elements in $H$. Let $\bar{H}$ be the set
of elements in $H$ which have no fixed points on $H^2$. We decompose
$\bar{H}$ into an  even $H_+$ and an odd $H_-$ pieces  defined by
the condition
\begin{equation}
    H_\pm = (\gamma\in \bar{H}: \ \ \chi (\gamma) =\pm 1 \}.
\end{equation}
Elements of $H_+$ are translations $\gamma$ which are  conformal
transformations (\ref{con}) with the property $a+d>2$.  For any
translation $\gamma$ there exist  a geodesic $L$ in $\Lambda$ which
is preserved by this translation. The reflection $q$ with respect to
$L$ commutes with $\gamma$. Thus $\Sigma_\gamma$ is generated by
$\gamma$ and $q$ which  fundamental domain is  (\ref{FUN}). The
contribution from even hyperbolic transformations to the trace
formula  is ( see (\ref{C}) )
\begin{equation}
   Z_+[h]=\frac{1}{4} \sum_{\gamma \in H_+} \frac{l_\gamma
g(\frac{l_\gamma}{2})}{\sinh \frac{l_\gamma}{2}}.
\end{equation}
Translations $\gamma^n$ for positive integers $n$ have the same
stability group $\Sigma_\gamma$ . Let $A_+$ be the set of primite
elements in $H_+$ ( all elements in $H_+$ can be obtained by
multiplications of elements in $A_+$ ). The even trace formula  can
be rewritten as
\begin{equation}\label{EVEN}
   Z_+[h]=\frac{1}{4} \sum_{\gamma \in A_+} \sum_{n=1}^\infty \frac{l_\gamma
g(\frac{nl_\gamma}{2})}{\sinh \frac{nl_\gamma}{2}}.
\end{equation}
\\
\\
Let now  $\gamma\in H_-$. A transformation  $\gamma^2$  is a
translation. Therefore  there exist a geodesic $L$ which is
preserved by  it. Let $q$ be the reflection with respect to $L$ and
$\gamma=q\gamma_0$. If we transport  $L$ into the imaginary axis
$L_0$ the translation $\gamma^2$ will be of the diagonal form
(\ref{trans}) with the length $l_{\gamma^2}$ and the reflection $q$
transforms into  $q_0$  given by (\ref{reflection}). Let
\begin{equation}
    \gamma_0= \left(
       \begin{array}{cc}
         a & b \\
         c & d \\
       \end{array}
     \right).
\end{equation}
The action of $q_0$ in the matrix form is
\begin{equation}
    q_0\left(
       \begin{array}{cc}
         a & b \\
         c & d \\
       \end{array}
     \right)q_0=
\left(
  \begin{array}{cc}
    a & -b \\
    -c & d \\
  \end{array}
\right).
\end{equation}
The condition
\begin{equation}
    \gamma^2=\left(
                \begin{array}{cc}
                  e^{\frac{l_{\gamma^2}}{2}} & 0 \\
                  0 & e^{-\frac{l_{\gamma^2}}{2}} \\
                \end{array}
              \right)
\end{equation}
implies that $ \gamma_0$ is of the diagonal form with $l_{\gamma_0}=
\frac{1}{2} l_{\gamma^2}$.  Thus  $\Sigma_\gamma$ is generated by $
\gamma_0$ and $q$ with the fundamental domain (\ref{FUN}) where now
$l_\gamma$ is defined to be the half of the length of the
translation $\gamma^2$. The contribution from odd hyperbolic
transformations to the trace formula is
\begin{equation}
   Z_-[h]=\frac{1}{4} \sum_{\gamma \in H_-} \frac{l_\gamma
g(\frac{l_\gamma}{2})}{\cosh \frac{l_\gamma}{2}}.
\end{equation}
Transformations  $\gamma^{2n+1}$ for positive integers $n$ have the
same stability group.  The odd trace formula takes the form
\begin{equation}\label{ODD}
   Z_-[h]=\frac{1}{4} \sum_{\gamma \in A_-} \sum_{n=1}^\infty \frac{l_\gamma
g((n+\frac{1}{2})l_\gamma )}{\cosh (n+\frac{1}{2})l_\gamma )}
\end{equation}
where  $A_-$ is  the set of primite elements in $H_-$.
\\
\\
Collecting all above terms we arrive at the Selberg's trace formula
in $X$
\begin{equation}\label{ST}
    Z[h]=Z_0[h]+Z_V[h]-Z_L[h]+Z_+[h]-Z_-[h]
\end{equation}
where $Z_0[h]$ given by (\ref{FREE}) is the trace over the spectra
of  the two dimensional hyperboloid $H^2$;  $Z_V[h]$ and $Z_L[h]$
given by (\ref{VERTEX}) and (\ref{GEODESIC}) includes effects of
vertexes and edges  which form the boundary of $X$; $Z_+[h]$ and
$Z_-[h]$ given by (\ref{EVEN}) and (\ref{ODD}) are contributions
from even and odd hyperbolic transformations. The Selberg trace
formula for hyperbolic groups which describe the spectra of oriented
manifolds without boundaries contains the free $Z_0$ and even $Z_+$
terms only \cite{Ran}. The remaining terms in (\ref{ST}) are related
to the boundary effects.
\\
\\
Before closing this Section we give the example of symmetric M-
polygon $X$ with equal edges  and angles between them. It  is
bounded by $M$ geodesics
\begin{equation}
    L_j= ( \tan ( \frac{\phi}{2} + (j-1) \frac{\pi }{M}),  \tan ( -\frac{\phi}{2} +(j-1) \frac{\pi }{M}
   ), \ \ \ j=1,2, \dots M
\end{equation}
where
\begin{equation}
    \sin\phi = \frac{\sin \frac{\pi}{M}}{\cos \frac{\pi}{P}}
\end{equation}
and $P$ is the period of the vertexes in $X$. The reflection with
respect to $L_j$ is
\begin{equation}
    q_j = k^{j-1} \gamma q k^{-(j-1)}
\end{equation}
where $\gamma$ is the diagonal matrix (\ref{trans}) with the length
given by
\begin{equation}
    \cosh l = \frac{\cos \frac{\pi}{P}}{\sin \frac{\pi}{M}},
\end{equation}
$q$ is the reflection with respect to the geodesic $(-1,1)$
\begin{equation}
    q z=\frac{1}{\bar{z}}
\end{equation}
and $k$ is the rotation of the order $M$. When $M\gg 1$ the length
of the translation $\gamma$ becomes much greater than one. For
symmetric $M$-polygon $X$ with sufficiently large  number of
vertexes we can neglect the even $Z_+[h]$ and the odd $Z_-[h]$ terms
in the Selberg's trace formula. As the  result we have
\begin{equation}
    Z[h]\simeq Z_0[h]+Z_V[h]+ Z_L[h], \ \ \ \ \ M\gg 1.
\end{equation}
For $P=4$  that is when the angles between geodesics are right ones
we have
\begin{equation}\label{ST1}
    Z[h]\simeq \frac{S}{4\pi}\int_{-\infty}^\infty d\rho \rho \tanh\pi\rho h(\rho )+
    \frac{M}{4} \int_0^\infty \frac{dy g(y)  }{\cosh y}+
    \frac{M \ln M}{4} g(0)
\end{equation}
If $M$ is even then one may glue the opposite edges of $X$ to get
the sphere with $g=\frac{M}{2}$ handles. By the Gauss-Bonnet theorem
the area of $X$ is $4\pi (g-1)$.

\vspace{2cm}\noindent {\bf 3. The Casimir energy in polygons }
\\
We consider a scalar field in the three dimensional Clifford-Klein
space-time
\begin{equation}\label{CK}
ds^2=dt^2-R^2\frac{(dx)^2+(dy)^2}{y^2}
\end{equation}
By  quantizing canonically  the scalar field in a polygon $X$  the
two point function  is given by ( $t>t^\prime$ ) formal expression
\begin{equation}
   \langle 0| \phi (x)\phi (x^\prime)|0\rangle=\sum_k \frac{e^{-i\omega_k (t-t^\prime -i\varepsilon )}}{2\omega_k}
   \Psi_k(\vec{x}) \Psi_k(\vec{x}^\prime)
\end{equation}
where
\begin{equation}
    \omega^2_n=\frac{1}{R} (\rho^2_n+\frac{1}{4})
\end{equation}
and $\{ \rho^2_n+\frac{1}{4}\}_{n=0}^\infty$ and $\{
\Psi_k(\vec{x})\}_{n=0}^\infty$ are the set of eigenvalues and
eigenfunctions of the Laplacian $-L$. As long as $\varepsilon$ is
kept different from zero, the two point function is well defined
quantity, which is (for $t>t^\prime$ ) related to the positive
frequency part of the Feynman propagator. The vacuum energy density
$E(\vec{x})$ can be obtained from this propagator by applying a bi-
differential operator and then by taking a coincidence limit
\cite{Bir}
\begin{equation}
   E(\vec{x})=\frac{1}{2}\lim_{(t^\prime,x^\prime,y^\prime ) \rightarrow (t,x,y)}[\partial_t\partial_{t^\prime}+
   y^2(\partial_x\partial_{x^\prime}+\partial_y\partial_{y^\prime})]\langle 0| \phi (x)\phi
   (x^\prime)|0\rangle.
\end{equation}
Integrating the energy density over $X$  we arrive at
\begin{equation}\label{Evar}
    E_\varepsilon=-\frac{1}{2R}\frac{\partial }{\partial \varepsilon}\sum_{n=0}^\infty
    e^{-\varepsilon\sqrt{\rho^2_n+\frac{1}{4}}}
\end{equation}
which is the regularization in which we are interested: The Casimir
energy is a finite part of $ E_\varepsilon$ in
$\varepsilon\rightarrow 0$ limit. For static space-times this
regularization is equivalent to another  regularization techniques
\cite{Guido}.
\\
\\
The Selberg's trace formula (\ref{ST}) gives the possibility to
calculate  the Casimir energy in an arbitrary  $M$-polygon $X$
\begin{equation}
    E_\varepsilon= -\frac{1}{2R}\frac{\partial }{\partial \varepsilon} Z[h]
\end{equation}
where
\begin{equation}
  h(\rho )= e^{-\varepsilon\sqrt{\rho^2+\frac{1}{4}}}.
\end{equation}
To establish  the dependence of the vacuum energy in $X$ on the
boundary configuration explicitly we consider the symmetric
$M$-polygon with sufficiently large number of vertexes which we
introduced in the previous Section. The Selberg's trace formula
(\ref{ST1}) implies
\begin{equation}
    E\simeq E_0+E_V+E_L.
\end{equation}
The free energy of the Clifford-Klein space time
\begin{equation}
    E_0= -\frac{1}{2R}\frac{\partial }{\partial \varepsilon} Z_0[h]
\end{equation}
is  negative \cite{Byt1}
\begin{equation}
    E_0= -\frac{S}{2\pi R}\int_0^\infty \rho d\rho
    \frac{\sqrt{\rho^2+\frac{1}{4}}}{1+e^{2\pi\rho}}-\frac{S}{96\pi R}
\end{equation}
where  $S$ is the area of $X$.
\\
\\
The effect of vertexes is given by
\begin{equation}
    E_V= - \frac{1}{2R}\frac{\partial }{\partial \varepsilon} Z_V[h]
\end{equation}
or
\begin{equation}\label{EV}
    E_V= - \frac{M}{8R} \int_0^\infty \frac{dy  }{\cosh y}\frac{\partial }{\partial
    \varepsilon} g_\varepsilon (y)
\end{equation}
where
\begin{equation}
    g_\varepsilon (y)=\frac{1}{\pi}\int_0^\infty d\rho \cos y\rho
    e^{-\varepsilon\sqrt{\rho^2+\frac{1}{4}}}.
\end{equation}
Using the Taylor expansion for $\cos y\rho$ and the integral
representation
\begin{equation}\label{Bessel}
    K_m(\frac{\varepsilon}{2})=\frac{\sqrt{\pi}} {\Gamma
    (m+\frac{1}{2})}\varepsilon^m \int_0^\infty dx x^{2m}\frac{e^{-\varepsilon\sqrt{\rho^2+\frac{1}{4}}}}{\sqrt{\rho^2+\frac{1}{4}}}
\end{equation}
for the modified Bessel function ( see page 959 of \cite{Grad} ) we
arrive at
\begin{equation}
    g_\varepsilon (y)=\frac{1}{\pi } \sum_{m=0}^\infty
    \frac{(-)^m}{m!}(\frac{y}{2})^{2m} \frac{\partial}{\partial
    \varepsilon} \frac{K_m(\frac{\varepsilon}{2})}{\varepsilon^m}
\end{equation}
The integral ( see page 349 of \cite{Grad} )
\begin{equation}
    \int_0^\infty \frac{y^{2m}}{\cosh y}=(\frac{\pi}{2})^{2m+1}\mid
    E_{2m}\mid
\end{equation}
allows us to rewrite (\ref{EV}) as
\begin{equation}
    E_V=\frac{M}{16R} \sum_{m=0}^\infty
    \frac{(-)^m \mid E_{2m}\mid}{m!}(\frac{\pi}{4})^{2m}
    \frac{\partial^2}{\partial
    \varepsilon^2} \frac{K_m(\frac{\varepsilon}{2})}{\varepsilon^m}.
\end{equation}
Here $E_m$ are the Euler's numbers which can be defined from the
expansion
\begin{equation}
   \frac{1}{\cosh z}=\sum_{n=0}^\infty \frac{E_n}{n!} z^n, \ \ \ \
   \mid z\mid < \frac{1}{2}.
\end{equation}
From  the series representation for the modified Bessel function (
see page 961 of \cite{Grad})
\begin{eqnarray}
    K_m(z)&=&\frac{1}{2}\sum_{k=0}^{m-1}(-)^m
    \frac{(m-k-1)!}{k!}(\frac{z}{2})^{2k-m}+ \nonumber \\
    &+&(-)^{m+1}\sum_{k=0}^\infty
    \frac{(\frac{z}{2})^{2k+m}}{k!(m+k)!}[\ln \frac{z}{2}-\frac{\Psi (k+1)+\Psi (m+k+1)}{2}]
\end{eqnarray}
we observe that only the second term at $k=1$ has finite part in
$\varepsilon\rightarrow 0$ limit:
\begin{equation}\label{regul}
    \frac{\partial^2}{\partial
    \varepsilon^2}
    \frac{K_m(\frac{\varepsilon}{2})}{\varepsilon^m}=(-)^{m+1}
    \frac{C_m}{4^{m+2}(m+1)!}
\end{equation}
where
\begin{equation}\label{const}
    C_m=2+2C+4\ln 2 -\sum_{k=1}^{m+1}\frac{1}{k}
\end{equation}
and  $C=0,577...$ is the Euler number.  Thus the Casimir effect due
to the vertexes is
\begin{equation}
    E_V=-\frac{M}{256R} \sum_{m=0}^\infty
    \frac{ \mid E_{2m}\mid}{m!(m+1)!}(\frac{\pi}{8})^{2m}C_m
\end{equation}
The coefficient $C_m$ becomes negative for $m\gg 1$. Since the terms
in the above series decrease fast we conclude that $E_V$ is
negative.
\\
\\
The effect of edges which form the boundary of $X$  is given by
\begin{equation}
    E_L=  \frac{1}{2R}\frac{\partial }{\partial \varepsilon} Z_L[h]
\end{equation}
or
\begin{equation}\label{EL}
    E_V=  \frac{M\ln M}{8R} \int_0^\infty \frac{dy  }{\cosh y}\frac{\partial }{\partial
    \varepsilon} g_\varepsilon (0)
\end{equation}
The integral representation (\ref{Bessel}) implies
\begin{equation}
    E_V=  -\frac{M\ln M}{8\pi R} \int_0^\infty \frac{dy  }{\cosh y}\frac{\partial^2 }{\partial
    \varepsilon^2} K_0(\frac{\varepsilon}{2})
\end{equation}
By the virtue of (\ref{regul}) we get the final result
\begin{equation}
    E_V=  \frac{C_0}{128\pi R}M\ln M
\end{equation}
which is positive. At $M\rightarrow \infty$ the contribution to the
Casimir energy  from the vertexes becomes much greater than the free
energy $E_0$ and effects from the edges of $X$. In this limit  we
have
\begin{equation}
    E\simeq \frac{C_0}{128\pi R}M\ln M.
\end{equation}
The Casimir energy for the symmetric $M$-polygon with sufficiently
large number of vertexes is positive and increases with the number
of vertexes. To see the rate of the increasing we  divide the above
expression on the area of $X$ which is for even M is given by
$S=2\pi M$ ( when $M\gg 1$ the area for any polygon $X$ is a linear
function of $M$ ). The vacuum energy per unit area increases as
logarithm of M.
\\
\\
Before closing this Section we shortly discuss the contribution to
the vacuum energy from the even $Z_+$ and the odd $Z_-$ terms in the
Selberg's trace formula. The Casimir effect from even hyperbolic
transformations in $\Gamma$ is
\begin{equation}
   E_+=-\frac{1}{32\pi R} \sum_{\gamma \in A_+} \sum_{n=1}^\infty \frac{
K_1(\frac{nl_\gamma}{4})}{n\sinh \frac{nl_\gamma}{2}}
\end{equation}
which appears to be negative, as $E_0$ and $E_V$.  The latter ones
are also related to even transformations in $\Gamma$ (
transformations which do not contain  a reflection ). Even elements
in the fundamental group give negative contribution to the vacuum
energy. If $\Gamma$ is strictly hyperbolic group then we have only
even transformations. As the  result the Casimir energy for oriented
two dimensional hyperbolic manifolds without boundaries is negative
\cite{Byt1, Floyd}.
\\
\\
From the other side the Casimir effect from odd hyperbolic
transformations
\begin{equation}
   E_-=\frac{1}{32\pi R} \sum_{\gamma \in A_-} \sum_{n=1}^\infty \frac{
K_1(\frac{(2n+1)nl_\gamma}{4})}{n\cosh \frac{(2n+1)l_\gamma}{2}}
\end{equation}
is positive. The vacuum energy $E_L$ from edges of $X$ is also due
to odd transformations in  $\Gamma$: Odd transformations give
positive contribution to the vacuum energy.
\\
\\
To establish some physical consequences of the above observation we
shortly discuss the relation between  the group  $\Gamma_X$ which
defines $X$ and the spectrum $\Pi_X$ of the Laplacian (\ref{lapl})
on $X$ . Let us assume that there exist some geodesic in $X$ such
that the reflection with respect to it is the symmetry
transformation of $X$. Let $X^\prime$ be the subspace of $X$ which
is obtained by the identification of opposite points in $X$. The
fundamental group of $X^\prime$ is "bigger" than the one of $X$:
$\Gamma_{X^\prime}$ is generated by elements of $\Gamma_X$ and by
the above reflection. From the other side the spectrum
$\Pi_{X^\prime}$ of $X^\prime$ is "smaller" than the one of $X$: One
has to impose addition conditions on $\Pi_X$ to arrive at
$\Pi_{X^\prime}$. We have the following duality
\begin{equation}
    \Gamma_X\subset \Gamma_{X^\prime} \Longrightarrow \Pi_{X^\prime} \subset
    \Pi_X
\end{equation}
between transformations which defines polygons in $H^2$ and spectra
of the Laplacian on these  polygons. Even and odd transformations
are responsible for those modes in the spectra which give negative
and  positive contributions to the vacuum energy. However no
explicit formulas for the spectra $\Sigma_X$ and the above duality
is known  for hyperbolic polygons.

\vspace{1cm}\noindent {\bf Acknowledgments}: The author thanks
Turkish Academy of Sciences (TUBA) for its support.

\vspace{1cm}\noindent {\bf Appendix}
\\
\\
The  function (\ref{Kfunction}) can be rewritten as
\begin{equation}
   K(z.w)= -\frac{1}{\pi \sqrt{2}} \int_d^\infty dy
    \frac{g^\prime (y) }{\sqrt{\cosh y -\cosh d}}
\end{equation}
or
\begin{equation}\label{int}
   \Phi  (\xi )=-\frac{1}{\pi} \int_\xi^\infty\frac{d\eta
   Q^\prime(\eta)}{\sqrt{\eta-\xi}}.
\end{equation}
where we introduced new variables
\begin{equation}
   \Psi (2(\cosh d- 1))=K(z.w)
\end{equation}
and
\begin{equation}\label{F1}
   Q (2(\cosh y- 1))=g(y).
\end{equation}
Here $d$ is the geodesic distance (\ref{dist}) and $g(y)$ is the
function given by (\ref{gfunction}).
\\
\\
The integral (\ref{int})   has the inverse
\begin{equation}\label{inv}
   Q  (\eta)= \int_{-\infty}^\infty d\xi \Phi (\eta+\xi^2).
\end{equation}
 In the new variable $\tau^2=\eta-\xi$ the expression (\ref{int})
reads
\begin{equation}
   \Phi  (\xi )=-\frac{1}{\pi} \int_{-\infty}^\infty d\tau \frac{d}{d\xi}
   Q(\xi+\tau^2).
\end{equation}
Using (\ref{inv}) we have
\begin{equation}
   \Phi  (\xi )=-\frac{1}{\pi} \int_{-\infty}^\infty d\tau d\sigma  \frac{d}{d\xi}
   \Phi(\xi+\tau^2+\sigma^2).
\end{equation}
which in polar coordinates reads
\begin{equation}
   \Phi  (\xi )=-2 \int_{0}^\infty rdr   \frac{d}{d\xi}
   \Phi(\xi+r^2)=- \int_{0}^\infty dx   \frac{d}{dx}
   \Phi(\xi+x)= \Phi  (\xi ).
\end{equation}
Thus we have shown that (\ref{inv}) is the inverse transform of
(\ref{int}).
\\
\\
A) Let $\frac{P}{2}$ be an odd number. Then the contribution from
$k^n$ to the trace formula is
\begin{equation}\label{A}
  I(k^n)=\int_{H^2/G_0} d\mu   K (k^nw, w)=\frac{2}{P}\int_{H^2} d\mu   K (k^nw, w)
\end{equation}
where $P$ is the period of a vertex, $G_0$ is  the rotational group
of the order $\frac{P}{2}$. The distance is
\begin{equation}
   \cosh d(k^nw,w )=1+\frac{\sin^2\frac{2\pi n}{P}}{2y^2}\mid 1+w^2\mid^2
\end{equation}
In the polar coordinates $w=e^{\alpha}e^{i\psi}$ we have
\begin{equation}
   I(k^n)=\frac{2}{P}\int_{-\infty}^\infty d\alpha \int_{-\infty}^\infty dz
   \Psi (4\sin^2\frac{2\pi n}{P}(\sinh^2\alpha +z^2\cosh^2\alpha ))
\end{equation}
where $z=\cot\psi$. By the virtue of (\ref{inv}) we get
\begin{equation}
    I(k^n)=\frac{1}{P\mid \sin \frac{2\pi n}{P}\mid }\int_{-\infty}^\infty \frac{d\alpha}{\cosh \alpha} Q
       (4\sin^2\frac{2\pi n}{P}\sinh^2\alpha  )
\end{equation}
or
\begin{equation}
   I(k^n)=\frac{1}{P}\int_{-\infty}^\infty \frac{dt}{\sin^2\frac{2\pi n}{P}+t^2} Q
       (4t^2  )
\end{equation}
where $t= \sin\frac{2\pi n}{P}\sinh\alpha$.  If we put $t=\sinh y$
then the above formula together with (\ref{F1}) implies
\begin{equation}
    I(k^n)=\frac{1}{P}\int_{-\infty}^\infty \frac{dy g(y)\cosh y}{\sin^2\frac{2\pi
    n}{P}+\sinh^2y}.
\end{equation}
\\
\\
B) The contribution from $\gamma\in H_+$ to the trace formula is
\begin{equation}\label{C}
    I(\gamma )=\frac{1}{2}\int_{1<\mid z\mid < e^{l_\gamma} } d\mu   K (\gamma w, w)
\end{equation}
where $\gamma$ is the translation (\ref{trans}) with the length
$l_\gamma$. The distance is
\begin{equation}
    \cosh d(\gamma w, w )=1+2\sinh^2\frac{l_\gamma}{2}\frac{ \mid w\mid^2}{y^2}
\end{equation}
In the polar coordinates we have
\begin{equation}
   I(\gamma ) = \frac{1}{2} \int_1^{e^{l_\gamma}}\frac{dr}{r}\int_{-\infty}^\infty dz \Psi
    (4\sinh^2\frac{l_\gamma}{2} (1+z^2))
\end{equation}
or
\begin{equation}
   I(\gamma )=\frac{l_\gamma}{4\sinh\frac{l_\gamma}{2}}
   g(\frac{l_\gamma}{2}).
\end{equation}
\\
\\
C) The contribution from $\gamma\in H_-$ to the trace formula is
\begin{equation}\label{B}
J(\gamma ) = \frac{1}{2}\int_{1<\mid z\mid < e^{l_\gamma}} d\mu K
(\gamma w, w).
\end{equation}
Here $\gamma=q_0\gamma_0$ with $q_0$ and $\gamma_0$ being the
reflection (\ref{reflection}) and the translation (\ref{trans}) with
the length  $l_\gamma$. In the polar coordinates  we have
\begin{equation}
    d(q_0\gamma_0 w, w )=1+2(\sinh^2\frac{l_\gamma}{2}+
    \cosh^2\frac{l_\gamma}{2}z^2 )
\end{equation}
The  formula (\ref{inv}) implies
\begin{equation}
J= \frac{l_\gamma g(\frac{l_\gamma}{2})}{4\cosh \frac{l_\gamma}{2}}
\end{equation}

\end{document}